\documentclass[letter]{aa}  
\usepackage{graphicx}
\usepackage{txfonts}
\begin{document}
   \title{Discovery of diffuse radio emission at the center of the
   most X-ray-luminous cluster RX J1347.5$-$1145}

   \subtitle{}

   \author{M. Gitti
          \inst{1}
          \and
          C. Ferrari\inst{2}
          \and 
          W. Domainko\inst{3}
          \and
          L. Feretti\inst{4}
          \and
          S. Schindler\inst{2} 
          }

   \offprints{M. Gitti}

   \institute{INAF - Osservatorio Astronomico di Bologna,
via Ranzani 1, 40127 Bologna, Italy\\
\email{myriam.gitti@oabo.inaf.it} 
\and Institut f{\"u}r Astro- und Teilchen Physik, Universit{\"a}t Innsbruck,
Technikerstra{\ss}e 25, 6020 Innsbruck, Austria\\
\and Max-Planck Institute for Nuclear Physics, 
Saupfercheckweg 1, 69117 Heidelberg, Germany\\
\and INAF - Istituto di Radioastronomia, 
via Gobetti 101, 40129 Bologna, Italy
}

   \date{Received ; accepted }

 
  \abstract 
  { We report on new VLA radio observations of the distant cluster RX
    J1347.5-1145, which is the most luminous in X-rays.  }
  { We aim at investigating the possible presence of diffuse and
    extended radio emission in this very peculiar system which shows
    both a massive cooling flow and merging signatures.  }
  { New low resolution ($\sim$ 18 arcsec) VLA radio observations of
    this cluster are combined with higher resolution ($\sim$ 2 arcsec)
    data available in the VLA archive.  }
  { We discover the presence of a diffuse and extended ($\sim$ 500
    kpc) radio source centered on the cluster, unrelated to the radio
    emission of the central AGN.  The properties of the radio source,
    in particular a) its occurrence at the center of a massive cooling
    flow cluster, b) its total size comparable to that of the cooling
    region, c) its agreement with the observational trend between
    radio luminosity and cooling flow power, indicate that RX
    J1347.5-1145 hosts a radio mini-halo. We suggest that the radio
    emission of this mini-halo, which is the most distant object of
    its class discovered up to now, is due to electron re-acceleration
    triggered by the central cooling flow.  However, we also note that
    the morphology of the diffuse radio emission shows an elongation
    coincident with the position of a hot subclump detected in X-rays,
    thus suggesting that additional energy for the electron
      re-acceleration might be provided by the submerger event. 
    }
   {}


   \maketitle


\section{Introduction}

Diffuse and extended synchrotron emission has been detected in a
number of galaxy clusters, revealing the presence of relativistic
electrons and magnetic fields diffused into the intra-cluster medium
(ICM). These radio sources, characterized by steep radio spectra and
sizes of $\sim$ 1 Mpc, are usually classified as ``radio halos'' when
permeating the cluster center, ``radio relics'' when located in the
cluster external regions (e.g., Feretti \& Giovannini 2007).  The
presence of both halos and relics in galaxy clusters have been
suggested to be strongly connected to the existence of cluster-cluster
mergers, which would provide the energy for re-accelerating the
electrons to ultra-relativistic energies (primary models; e.g.,
Brunetti et al. 2001). Relativistic electrons could also result from
inelastic nuclear collisions between relativistic protons and thermal
ions of the ICM (secondary models; e.g. Blasi \& Colafrancesco 1999).
Up to now, the comparison of observational and analytical results is
in favor of primary models related to cluster mergers (e.g., Brunetti
2003, Blasi et al. 2007).  In particular, radio halos and relics are
not commonly seen in clusters characterized by short central cooling
time and high central brightness peak, the so-called cooling flow
clusters (e.g., Fabian 1994, Peterson \& Fabian 2006), since the
cooling flow is likely suppressed or disrupted by a major merger
(Gomez et al. 1997).

The central regions of cooling flow clusters host an exceptional class
of diffuse radio sources, called ``mini-halos'' due to their smaller
extension ($\sim$500 kpc). They usually surround a powerful radio
galaxy. However, their radio emission is not due to radio lobes fed by
an active galactic nucleus (AGN), but, again, to the presence of
relativistic particles and magnetic field in the ICM.  Gitti et
al. (2002) proposed that the synchrotron emission from radio
mini-halos could be due to a relic population of (primary)
relativistic electrons re-accelerated by MHD turbulence, with the
necessary energy supplied by the cooling flow process itself.
Alternatively, Pfrommer \& En{\ss}lin (2004) discussed the possibility
of radio emission from secondary electrons.  The successful
application of the re-acceleration model to two cooling flow clusters
(Gitti et al. 2002, 2004) has given support to a primary origin of the
relativistic electrons radiating in radio mini-halos, although
detailed observations of a large sample are required to understand the
origin of these sources.

RX J1347.5$-$1145 is the most X-ray-luminous cluster discovered to
date ($L_X = 6 \times 10^{45}$ erg s$^{-1}$ in the [2-10] keV energy
range, Schindler et al. 1995; Gitti \& Schindler 2004).  It shows a
very peaked surface brightness profile and hosts a massive cooling
flow in its center (Gitti \& Schindler 2004).  Sub-millimeter and
millimeter observations in its direction detected significant
Sunyaev-Zeldovich (SZ) signal and revealed complex morphological and
temperature structures of the cluster region (Komatsu et al. 1999,
2001; Pointecouteau et al. 1999, 2001). Recent \textit{Chandra} (Allen
et al. 2002) and \textit{XMM--Newton} (Gitti \& Schindler 2004)
observations identified a region with hot, bright X-ray emission
located $\sim 20$ arcsec to the main X-ray surface brightness peak in
the south-east direction, at a position coincident with the region of
enhanced SZ effect (Komatsu et al. 2001).  These results were
interpreted as indications of a subcluster merger, making RX
J1347.5$-$1145 a very peculiar system with a complex dynamical
evolution.  If on the one hand there is indication of a disturbed
dynamical state in the south-east region, on the other hand the strong
cooling flow points to a relatively long interval of time in which the
cluster has evolved undisturbed to a relaxed state (Gitti et
al. 2007).  This cluster is thus perfectly suited to study the
interplay of massive cooling flows and subcluster mergers, which are
very rarely observed in the same system, on the re-acceleration of
relativistic electrons.  The radio emission in the central 3$\times$3
arcmin$^2$ of RX J1347.5$-$1145, as seen from the NRAO VLA Sky Survey
(NVSS; Condon et al. 1998), shows a strong central source and some
hints of a possible extended emission.  In this Letter we present the
results from new high sensitivity 1.4 GHz radio observations of this
cluster, which led to the discovery of diffuse radio emission that may
be classified as a mini-halo.

RX J1347.5$-$1145 (hereafter RX J1347) is at a redshift of 0.451
\footnote{ With a Hubble constant of $H_0 = 70 \mbox{ km s}^{-1}
  \mbox{ Mpc}^{-1}$, and $\Omega_M = 1-\Omega_{\Lambda} = 0.3$, the
  luminosity distance is 2506 Mpc and the angular scale is 5.77 kpc
  arcsec$^{-1}$}, thus being the most distant cluster in which a radio
mini-halo has been detected.


\section{Observations and Data Reduction}

\begin{figure} 
\centering
\includegraphics[scale=0.42, angle=-90]{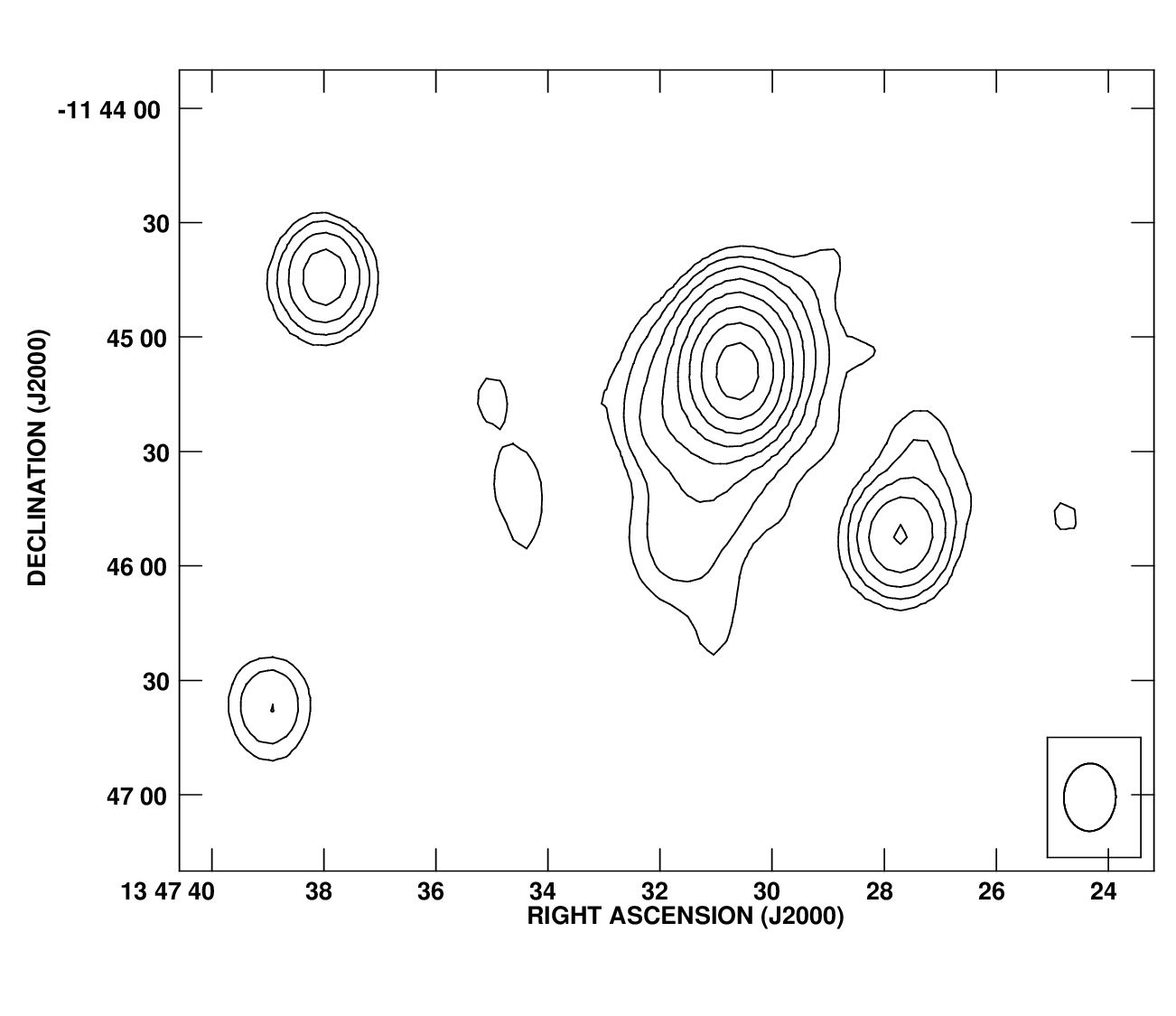}
\caption{
1.4 GHz VLA map of RX J1347 at a resolution of $17.8'' \times 13.6''$ 
(the beam is shown in the lower-right corner).
The r.m.s. noise is 0.04 mJy/beam. 
The first contour corresponds to 5$\sigma$, the ratio 
between two consecutive contours being 2.
}
\label{mappa-C.fig}
\end{figure}

\begin{figure} 
\centering
\includegraphics[scale=0.42, angle=-90]{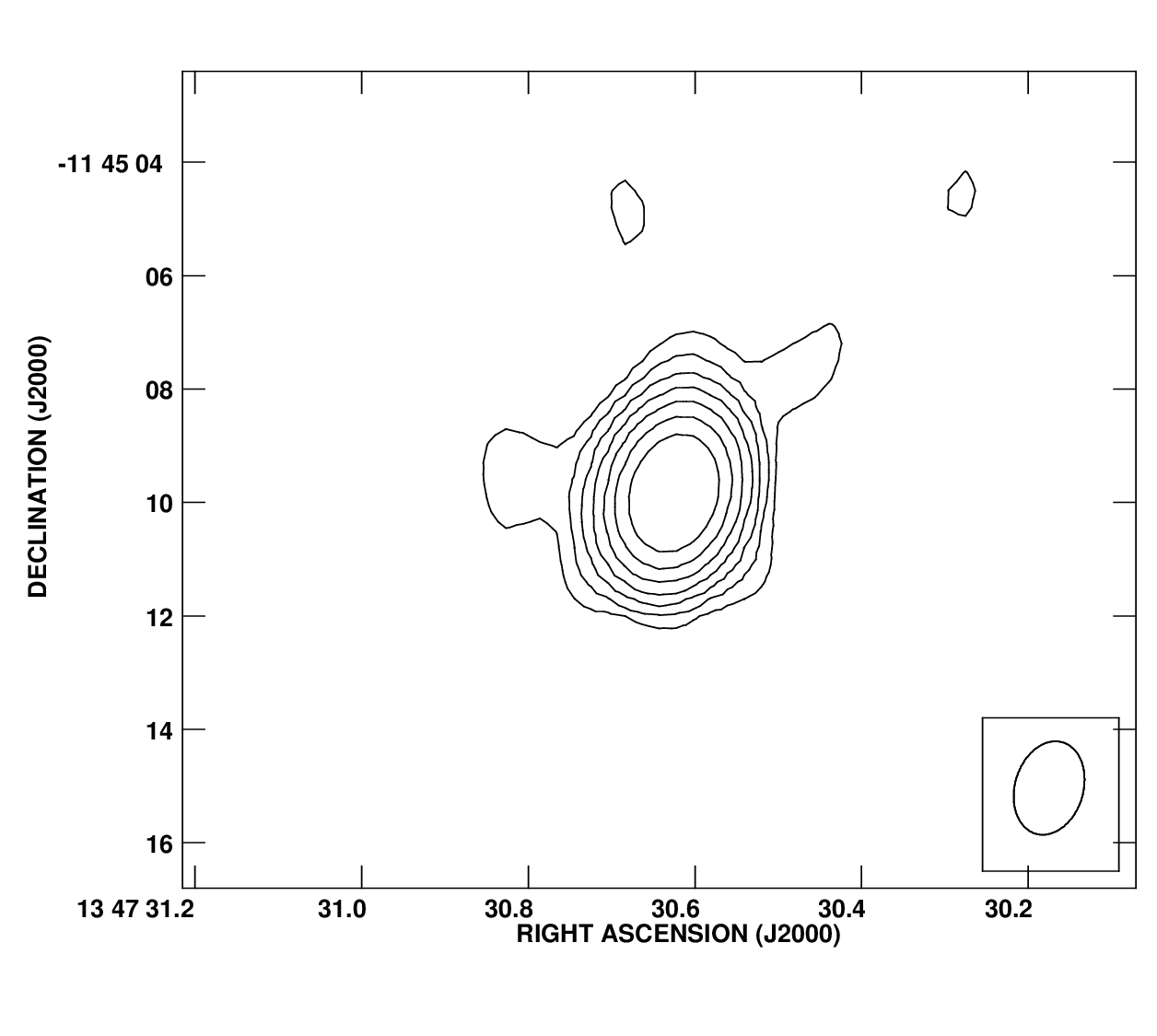}
\caption{
1.4 GHz VLA map of RX J1347 at a resolution of $1.7'' \times 1.2''$ 
(the beam is shown in the lower-right corner).
The r.m.s. noise is 0.03 mJy/beam.  
The first contour corresponds to 5$\sigma$, the ratio 
between two consecutive contours being 2.
}
\label{mappa-A.fig}
\end{figure}

We performed Very Large Array\footnote{The Very Large Array (VLA) is a
  facility of the National Radio Astronomy Observatory (NRAO).  The
  NRAO is a facility of the National Science Foundation, operated
  under cooperative agreement by Associated Universities, Inc.}
observations of the radio source RX J1347 at 1.4 GHz in
C-configuration, that we combined with same and higher resolution (VLA
in A-configuration) observations at the same frequency available in
the VLA archive (see Table \ref{vladata.tab} for details regarding
these observations).  In all observations the source 3C 286 is used as
the primary flux density calibrator, while the source 1351$-$148 is
used as the secondary phase calibrator.  Data reduction is done using
the NRAO AIPS (Astronomical Image Processing System) package.
Accurate editing of the uv data is applied to identify and remove bad
data.  Images are produced by following the standard procedures:
Calibration, Fourier-Transform, Clean and Restore.  Self-calibration
is applied to remove residual phase variations.  Data in array C from
the two different observations are reduced separately, in order to
analyze the possible existence of spurious features, and the resulting
images are then combined.  The final images, produced using the AIPS
task IMAGR, show the contours of the total intensity
(Figs. \ref{mappa-C.fig} and \ref{mappa-A.fig}).

\begin{table}
\begin{center}
\caption{VLA Data}
\hskip -.2truein
\begin{tabular}{cccccc}
\hline
\hline
Observation & Frequency & Bandwidth & Array & Exp. Time\\
Date & (MHz) & (MHz) & ~ & (h) \\
\hline
~&~&~&~\\
Sep-2005 & 1365/1435 & 50.0 & C & 2.0\\
Dec-1998 & 1385/1465 & 25.0 & C & 2.0\\
Apr-1998 & 1385/1465 & 25.0 & A & 3.0\\
\hline
\label{vladata.tab}
\end{tabular}
\end{center}
\end{table}

\section{Results}

Figure \ref{mappa-C.fig} shows the radio map of RX J1347 observed at
1.4 GHz with the VLA in C configuration, with a restoring beam of
$17.8'' \times 13.6''$.  The source is clearly extended with a strong
central component and an amorphous morphology elongated toward the
south (S). The total size is $\sim 107''$ (620 kpc) in the north-south
(NS) direction and the total flux density is $55.3 \pm 0.6$ mJy (see
Table \ref{risradio.tab}).  The source visible to the south-west (SW)
of the extended emission is associated to a cluster galaxy, and has a
total flux density of $4.50 \pm 0.06$ mJy.

Figure \ref{mappa-A.fig} shows the radio map of the very central field
of RX J1347 observed at 1.4 GHz with the VLA in A configuration, with
a restoring beam of $1.7'' \times 1.2''$.  With this high resolution
map it is possible to image the central source and to spot the
presence of discrete sources that can contribute to the total flux and
extended morphology detected at lower resolution.  The central source,
located at RA(J2000):$13^{\rm h} 47^{\rm m} 30.6^{\rm s}$,
Dec(J2000):$-11^{\circ} 45' 09.9''$, is resolved, showing hints of
faint structures (detected at a 6$\sigma$ level) emanating to the east
(E) and north-west (NW) directions from the center. No relevant
structures are visible toward the S direction.  The total flux density
is $29.8 \pm 0.3$ mJy.  The high resolution structure of the central
source strongly favours the possibility that it is unrelated to the
diffuse emission detected at lower resolution.  In this high
resolution map the source to the SW, located at RA(J2000):$13^{\rm h}
47^{\rm m} 27.7^{\rm s}$, Dec(J2000):$-11^{\circ} 45' 53.4''$ (not
visible in Fig. \ref{mappa-A.fig}), has a total flux density of $1.90
\pm 0.04$ mJy.  Its morphology suggests that it is a head-tail radio
galaxy. The difference in the fluxes measured at high and low
resolution is thus due to the diffuse emission associated to the tail
of the radio source.

\begin{figure} 
\centering
\includegraphics[scale=0.45]{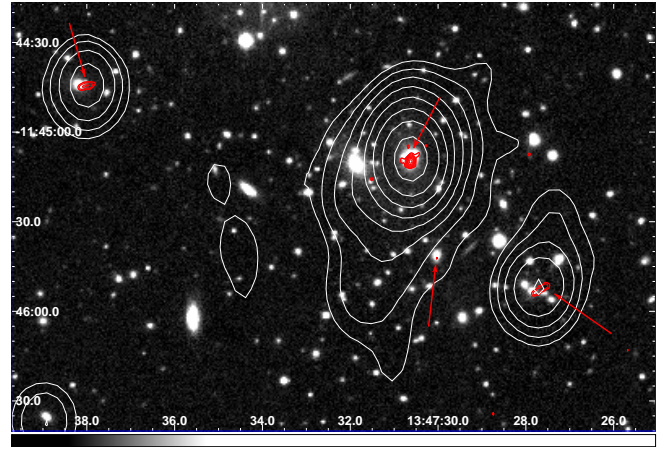}
\caption{
Optical I-band image of the central field of RX J1347.
Overlayed are the 1.4 GHz radio contours at 
low \textit{(white)} and high \textit{(red)} resolution.   
The beam, r.m.s. noise and contours are the same as in Figs. 
\ref{mappa-C.fig} and \ref{mappa-A.fig}, respectively.
The arrows show the discrete radio sources identified for the subtraction
from the low resolution radio map. 
}
\label{radio-ottico.fig}
\end{figure}

\begin{figure} 
\centering
\includegraphics[scale=0.45]{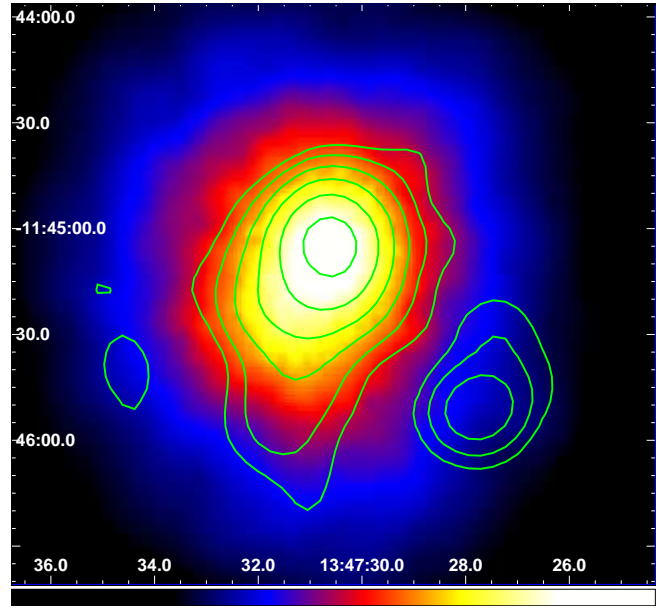}
\caption{
Adaptively smoothed \textit{XMM--Newton} image 
of the central region of RX J1347 (data from Gitti \& Schindler 2004).
A bright subclump is visible $\sim 20''$ to the SE of the center 
(see also Allen et al. 2002).
Overlayed are the 1.4 GHz contours of the diffuse radio emission 
at a resolution of $18'' \times 18''$ after the subtraction
of the discrete radio sources identified at higher resolution
(see Fig. \ref{radio-ottico.fig}).
The r.m.s. noise is 0.05 mJy/beam. 
The first contour corresponds to 5$\sigma$, the ratio 
between two consecutive contours being 2.
}
\label{radio-X.fig}
\end{figure}

\begin{figure} 
\centering
\includegraphics[scale=0.45]{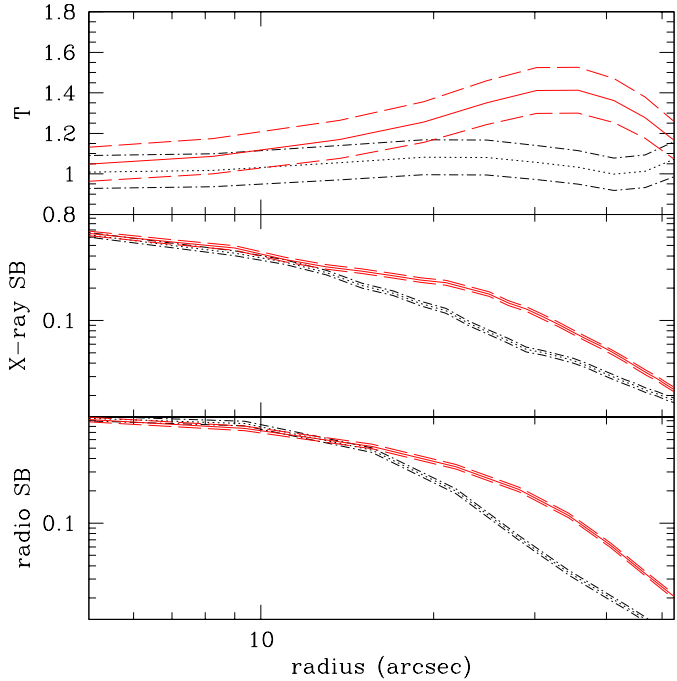}
\caption{ The radial profiles extracted along the SE direction
  including the subclump ($40^{\circ}$ east of south), which are shown
  with the solid lines, are compared to the radial profiles
  extracted along the opposite direction ($40^{\circ}$ west of north),
  shown with the dotted lines.  The profiles
  (embedded in the 1$\sigma$ error envelopes shown with thin
    dashed and dot-dashed lines, respectively) are normalized with
  respect to the central values.  Clear enhancements are visible in
  all profiles along the SE direction.  
  \textit{Upper panel}:
  Temperature profile derived from the \textit{XMM-Newton} hardness
  ratio map presented by Gitti \& Schindler (2004, Fig. 4).
  \textit{Middle panel}: 
  X-ray surface brightness profile derived from \textit{XMM-Newton} 
  data (Gitti \& Schindler 2004).  
  \textit{Lower panel}: 
  Radio surface brightness profile of the mini-halo derived
  from the map shown in Fig. \ref{radio-ottico.fig}.  }
\label{cfr-profiles.fig}%
\end{figure}

In order to put in evidence the diffuse emission, it is essential to
subtract the contribution of discrete radio sources.  In a
conservative way, they are identified as the radio sources detected in
the high resolution map at $5\sigma$ level, having obvious association
with an optical galaxy.  The four sources selected with this method
are indicated by red arrows in Fig. \ref{radio-ottico.fig}.  The clean
components of the discrete sources are restored with a circular beam
of $18''$ and then subtracted from the low resolution map convolved
with the same beam.  This procedure allows us to estimate the total
flux density of the diffuse radio emission ($S_{\rm 1.4GHz}= 25.2 \pm
0.3$ mJy) and derive its surface brightness map, which is shown in
Fig. \ref{radio-X.fig} overlayed onto the X-ray image.  The size of
the diffuse radio emission ($\sim 435 \times 600$ kpc) is comparable
to that of the cooling region ($\sim 420$ kpc, Gitti et al. 2007),
thus indicating that the diffuse radio source may be classified as a
mini-halo.  We also note that the morphology of the radio emission
shows an elongation coincident with the position of the hot SE
subclump detected in X-rays (Allen et al. 2002; Gitti \& Schindler
2004).  Additionally, the SE excess in the ICM X-ray brightness and
temperature radial profiles corresponds to a clear excess in the radio
surface brightness of the diffuse radio source (see
Fig. \ref{cfr-profiles.fig}). The excess X-ray bolometric luminosity
of the SE quadrant between radii of $\sim 10''$ and $\sim 35''$, where
the subclump is most apparent at the \textit{Chandra} resolution,
corresponds to $\sim 5$ \% of the total cluster luminosity (Allen et
al. 2002).  Interestingly, we estimated that the excess radio flux of
the SE quadrant in the region where the enhancement is most evident in
the radio surface brightness profile (see Fig. \ref{cfr-profiles.fig},
lower panel) corresponds to $\sim 5$ \% of the total radio flux
density of the mini-halo.

The main radio results concerning the central source are summarized in
Table \ref{risradio.tab}, where we also report the monochromatic radio
power at each frequency calculated as $ P_{\nu}=4 \pi \, D_{\rm L}^2
\, S_{\nu} $ (here $D_{\rm L}$ is the luminosity distance and
$S_{\nu}$ is the flux density at the frequency $\nu$).

\begin{table*}
\caption{
\label{risradio.tab}
Radio results for RX J1347.
The size is estimated in the EW-NS directions along the maximum extent 
of the source.
}
\centerline{
\begin{tabular}{ccccccc}
\hline
\hline
Map & Beam & rms & Size & Peak & Total flux & Radio Power\\
~& (\arcsec$^2$)  & (mJy/beam) & (\arcsec$^2$) & (mJy/beam) & (mJy)& ($10^{24}$W Hz$^{-1}$)\\
\hline
~&~&~&~&~&~&~\\
array C & 17.8$\times$13.6 & 0.04 & 73.0$\times$107.4 & 37.8 (0.4) & 55.3 (0.6) & 41.5 (0.5)\\
array A & 1.7$\times$1.2 & 0.03 & 6.8$\times$5.1 & 27.6 (0.3) & 29.8 (0.3) & 22.4 (0.2)\\
subtracted & 18.0$\times$18.0 & 0.05 & 75.6$\times$103.7 & 11.0 (0.1) & 25.2 (0.3) & 18.9 (0.2) \\
\hline
\end{tabular}
} 
\end{table*}

\section{Discussion and Conclusions}

As already discussed in the Introduction, RX J1347 is a peculiar
cluster.  It hosts a massive cooling flow and shows a disturbed
dynamical state in the SE quadrant. Furthermore, it is dominated by
two Brightest Cluster Galaxies (BCGs).  If on the one hand this is
unusual for massive cooling flow clusters, on the other hand based on
its optical spectrum the western BCG is an AGN, with typical emission
lines of giant ellipticals at the center of cooling flow clusters
(Cohen \& Kneib 2002). The presence of two dominant galaxies is
probably related to a submerger event, since the giant ellipticals
observed in the center of the cluster could be the BCGs of the two
colliding subclusters.  Indications of a recent or ongoing submerger
event come also from the detection of a hot X-ray subclump in the SE
quadrant (Allen et al. 2002; Gitti \& Schindler 2004).  Interestingly,
we detect diffuse radio emission at the cluster center.  Its
occurrence in a massive cooling flow cluster and its total size
comparable to that of the cooling region ($\sim 420$ kpc, Gitti et
al. 2007) indicate that the diffuse radio source may be classified as
a mini-halo.

Gitti et al. (2002) developed a theoretical model which accounts for
the origin of radio mini--halos as related to electron re-acceleration
by MHD turbulence in cooling flows.  In this model, the necessary
energetics to power radio mini-halos is supplied by cooling flows
themselves, through the compressional work done on the ICM and the
frozen-in magnetic field. This supports a direct connection between
cooling flows and radio mini-halos.  A full application of the model
would be necessary to test its consistency with the observational
properties of RX J1347, as done for the mini-halos in Perseus (Gitti
et al. 2002), and A2626 (Gitti et al. 2004).  Unfortunately, this is
not possible with the present data because information on the spectral
index properties, essential to test two out of the three observational
constraints predicted by the model (namely, the brightness profile,
the total radio spectrum and the radial spectral steepening, Gitti et
al. 2002), is missing.  Here, we can test qualitatively the
consistency of the observational X-ray and radio properties of RX
J1347 with some correlation predicted by the model.  In particular,
since the re-acceleration model assumes a connection between the
origin of the synchrotron emission and the cooling flow, a trend
between the radio power on the mini-halos and the maximum power of the
cooling flows is expected. This has been observed in a first sample of
mini-halos selected by Gitti et al. (2004).  By plotting the
integrated radio luminosity $\nu P_{\nu}$ versus the maximum power of
the cooling flow $P_{\rm CF} = \dot{M} k T/\mu m_{\rm p}$ (here
$\dot{M}$ is the mass accretion rate, $k$ the Boltzmann constant, $T$
the ICM temperature, $\mu \approx 0.61$ the molecular weight, and
$m_p$ the proton mass), these authors found that the strongest radio
mini--halos are associated with the most powerful cooling flows.  In
more convenient units, the expression for the cooling flow power can
be rewritten as: $P_{\rm CF} \sim 10^{41} (\dot{M}/1 M_{\odot} \,{\rm
  yr}^{-1})(k T/1 \,{\rm keV}) \,{\rm erg \,s}^{-1}$.  We estimate
this quantity from the \textit{XMM-Newton} data presented by Gitti \&
Schindler (2004). To be consistent with the mini-halo sample data
collected from literature (Gitti et al. 2004), we adopt the overall
cluster temperature $k T = 10.0 \pm 0.3$ keV (Gitti \& Schindler 2004)
and mass accretion rate $\dot{M} = 1600^{+140}_{-650} \,M_{\odot}
\,{\rm yr}^{-1}$, derived by fitting the central spectrum with a
standard cooling flow model.

\begin{figure} 
\centering
\includegraphics[scale=0.45]{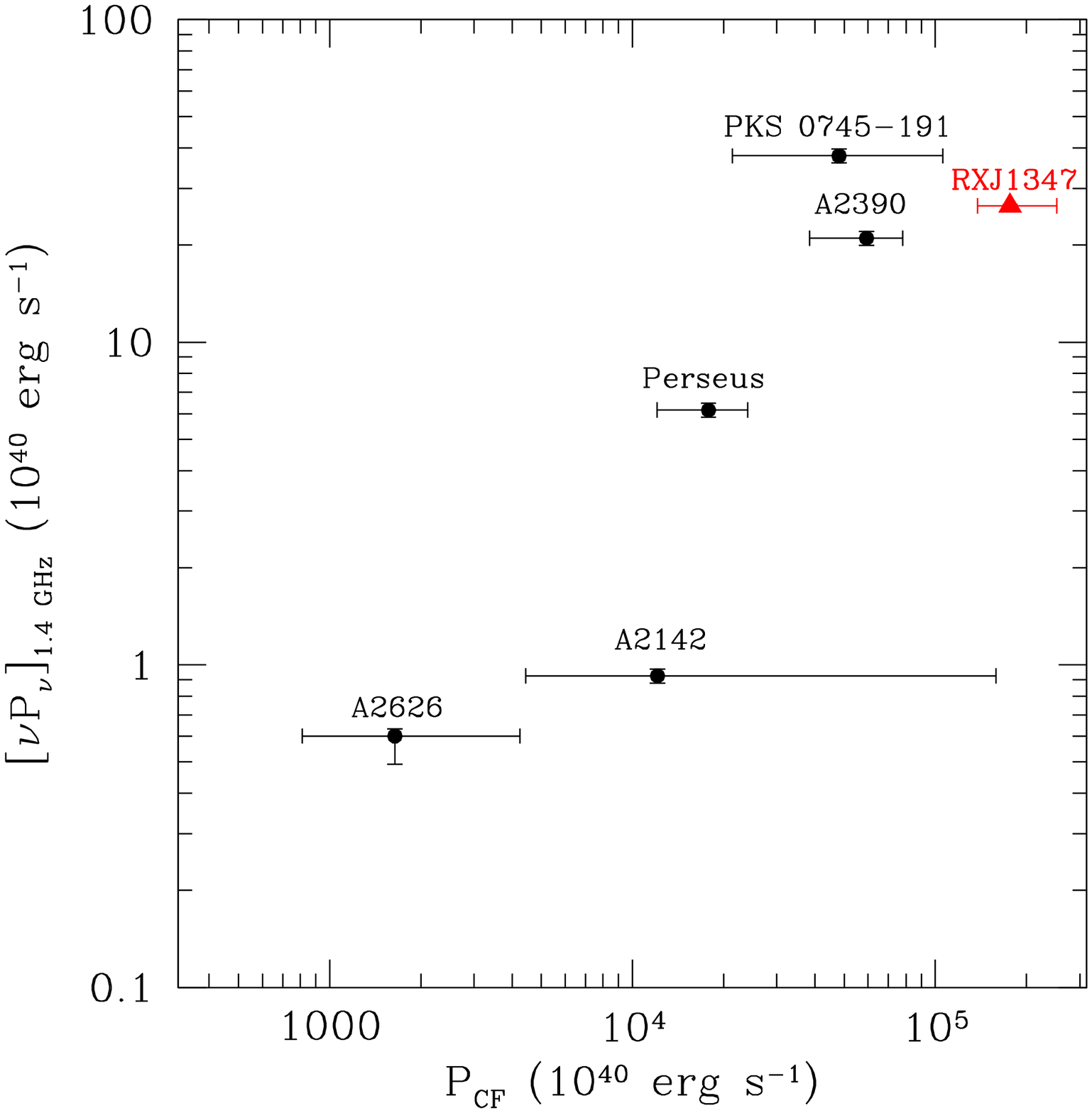}
\caption{
Integrated radio power at 1.4 GHz, $\left[ \nu P_{\nu} \right]_{\rm 1.4 GHz}$,
vs. cooling flow power, $P_{\rm CF} = \dot{M} k T/\mu m_{\rm p}$, 
for the mini--halo clusters in Gitti et al. (2004).
The triangle represents RX J1347 values estimated in this work
($P_{\rm CF}$ from \textit{XMM-Newton} data taken from
Gitti \& Schindler 2004).
}
\label{radio-cf.fig}
\end{figure}

In Figure \ref{radio-cf.fig} we show with a triangle the radio
power $\left[ \nu P_{\nu} \right]_{\rm 1.4 GHz} = (2.65 \pm 0.03)
\times 10^{41} \,{\rm erg \,s}^{-1}$ and cooling flow power $P_{\rm
  CF} = 1.6^{+0.2}_{-0.7} \times 10^{45} \,{\rm erg \,s}^{-1}$ of RX
J1347 estimated in this work, overlayed onto the values measured for
the mini-halo sample (Gitti et al. 2004).  We note that RX J1347
follows nicely the observed trend, resulting one of the most powerful
objects in the sample.  This gives support to the possible
classification of its diffuse radio source as a mini-halo triggered by
the central cooling flow.  However, the excess and elongation of the
radio surface brightness coincident with the position of the hot
subclump detected in X-rays (Figs.  \ref{radio-X.fig} and
\ref{cfr-profiles.fig}) might indicate that, besides the energy
supplied by the central cooling flow, additional energy for the
electron re-acceleration could be provided by the submerger. From this
analysis we thus get indications that cluster mergers and cooling
flows may act simultaneously in powering diffuse radio emission in the
rare and peculiar clusters in which they coexist.

Further theoretical and observational studies are currently in
progress and will give a more definite answer regarding the nature of
the diffuse emission in RX J1347.  In a more general context, we are
also carrying out a project aimed at selecting and studying a large
sample of radio mini-halos in order to further investigate their
origin and their connection with giant radio halos.

\begin{acknowledgements}
  We thank the anonymous referee for useful comments.
  We thank T. Erben for providing the optical image of the cluster. MG
  thanks D. Dallacasa for his helpful advises during the radio data
  analysis.  CF acknowledges financial support of Austrian Science
  Foundation (FWF) through grant number P18523, and
  TirolerWissenschaftsfonds (TWF) through grant number UNI-0404/156.
\end{acknowledgements}

\end{document}